\documentclass{aa}
\usepackage{amsmath,graphicx,amssymb}
\usepackage[varg]{txfonts}
\usepackage{natbib}
\usepackage{pst-eps}
\newcommand{\zav}[1]{\left(#1\right)}

\newlength\staretab

\newcommand\de{\text{d}}

\begin{document}

\title{Stochastic light variations in hot stars from wind instability: Finding photometric signatures and testing against the TESS data}

\author{J.~Krti\v{c}ka\inst{1} \and A.~Feldmeier\inst{2}}

\institute{Department of Theoretical Physics and Astrophysics, Faculty of
           Science, Masaryk University, Kotl\'a\v rsk\' a 2, Brno,
           Czech Republic \and
           Institut f\"ur Physik und Astronomie, Universit\"at Potsdam,
           Karl-Liebknecht-Stra{\ss}e 24/25, Potsdam-Golm, Germany}

\date{Received}

\abstract{Line-driven wind instability is expected to cause small-scale wind
inhomogeneities, X-ray emission, and wind line profile variability. The
instability can already develop  around the sonic point if it is initiated close
to the photosphere due to stochastic turbulent motions. In such cases, it may
leave its imprint on the light curve as a result of wind blanketing.}{We study
the photometric signatures of the line-driven wind instability.}{We used
line-driven wind instability simulations to determine the wind variability close
to the star. We applied two types of boundary perturbations: a sinusoidal one that enables us to study in detail the development of the instability and a
stochastic one given by a Langevin process that provides a more realistic
boundary perturbation. We estimated the photometric variability from the
resulting mass-flux variations. The variability was simulated assuming that the
wind consists of a large number of independent conical wind sectors. We compared the
simulated light curves with TESS light curves of OB stars that show
stochastic variability.}{We find two typical signatures of line-driven wind
instability in photometric data: a knee in the power spectrum of magnitude
fluctuations, which appears due to engulfment of small-scale structure by larger
structures, and a negative skewness of the distribution of fluctuations, which is
the result of spatial dominance of rarefied regions. These features endure even
when combining the light curves from independent wind sectors.}{The stochastic
photometric variability of OB stars bears certain signatures of the line-driven
wind instability. The distribution function of observed photometric data shows
negative skewness and the power spectra of a fraction of light curves exhibit a knee.
This can be explained as a result of the line-driven wind instability triggered
by stochastic base perturbations.}


\titlerunning{Stochastic light variations in hot stars due to wind instability}

\authorrunning{J.~Krti\v{c}ka \& A.~Feldmeier}
\maketitle

\section{Introduction}

The stellar winds of hot stars are mostly driven by radiative acceleration in
the lines of such elements as carbon, oxygen, silicon, and iron \citep{lusol,cak}.
This acceleration is inherently unstable. Due to the Doppler effect, small
positive perturbations of the velocity displace the lines from their equilibrium
positions and expose the line to unabsorbed stellar flux, which leads to
a substantial increase in the radiative force \citep{MacGregor,karl,ornest}.

The perturbations from the line-driven wind instability steepen into reverse
shocks \citep{ocr}, providing an explanation for the generic X-ray emission observed
in hot stars \citep{felpulpal}. In addition, the instability may magnify the
atmospheric density perturbations that exist due to subsurface convection or
stellar oscillations \citep{kant,ae,jian}, leading to a complex wind structure
typically referred to as clumping. Therefore,  one single process may, in fact, connect
the photospheric line profile variations \citep{khos,marmarvar,aero}, macroturbulent
broadening \citep{dufbvele,markopulos}, and wind variability.

Temporal variability on the part of the wind mass-flux is another consequence of the
line-driven wind instability. This contributes to photometric variations of hot
stars attributed to a process called wind blanketing. Wind blanketing refers to the
absorption of photospheric radiation by the wind and its partial reemission back
into the stellar photosphere \citep{acko}. As the atmosphere becomes more
opaque, the temperature gradient steepens to transfer the same flux through the
opaque region. Effectively, this leads to the heating of the photosphere.
Therefore, the inclusion of wind blanketing is important for the determination
of atmospheric properties of hot stars with winds \citep{boh,prvnifosfor,okali}.
With the increasing opacity, photons progressively escape into spectral regions
where the gas is more transparent, which leads to flux redistribution.
Therefore, as a result of wind blanketing, a small fraction of the radiative
flux emitted by hot stars is absorbed by their winds and redistributed towards
longer wavelengths. The amount of redistributed flux depends on the mass-loss
rate. Consequently, any process that causes either a temporal or lateral
variability of the wind in the regions just above the hydrostatic photosphere also
 inevitably  leads to photometric variability. There are several such effects
that can cause wind variability, including magnetic fields \citep{udo}, bright
surface spots \citep{duo}, and  line-driven instability
\citep{ocr,felpulpal}.

\citet{oblavar} used the dependence of the redistributed flux on the mass-loss
rate from global wind models \citep{magvar} and the instability simulations of
\citet{felpulpal} to predict the light variability of hot stars. They showed
that the combined effects of wind blanketing and instability lead to a
stochastic light variability with amplitude on the order of millimagnitudes.

This effect can be compared to the stochastic variability observed in hot stars.
Stochastic variability has a relatively low amplitude and although, in
some cases,  it can
be  observed from the ground \citep[e.g.][]{kourbos}, for most stars, precise
satellite photometry is required for it to be detected \citep{blomcor,ram,simondia}.
Typically, stochastic variability is attributed to surface oscillations driven
by convective motions \citep{aero,asid,leco}.

In the present paper, we use the approach of \citet{felto} to
better understand the signature of perturbations in the photometry and its
dependence on the inner boundary conditions. We study the dependence of the
light variability at the height where the optical flux originates and
we determine the photometric signatures of the line-driven wind instability. 

\section{Line-driven wind instability simulations}

A consistent treatment of the wind blanketing variability due to the line-driven
wind instability is a formidable problem. A suitable approach would require coupling 3D
radiative hydrodynamic simulations with a detailed solution of the radiative
transfer equation and the determination of nonequilibrium level populations. To
make the problem tractable, we split it into two parts: the simulation of the
line-driven wind instability and subsequent calculation of light variability
based on these hydrodynamic simulations. Our calculations are done in 1D and we
construct 3D wind models by dividing the visible stellar surface into patches
and integrating over the whole surface under the assumption that individual
patches are independent.

\subsection{Hydrodynamic equations and their solution}
\label{hydroeq}

The present level of sophistication of time-dependent numerical wind
simulations including the line-driven instability (LDI) is defined in
the following papers: \citet{owpu0,owpu}, introducing a nonlocal
escape-integral source function (EISF) for the scattered radiation
field;
\citet{sunowroz}, demonstrating the transition of
the solution topology near the sonic point from X-type to nodal type
in dependence on the strength of the diffuse radiation field and the
ratio of thermal-to-sound speed; and
\citet{sundsim}, performing
two-dimensional planar simulations of the LDI with the smooth source
function (SSF) approach of Owocki (1991) for the scattered radiation
field.

The present paper takes a complementary approach and calculates the
radiative line force in the simplest possible setting that still
shows the LDI, with force-per-mass given by
\begin{equation}
\label{grad}
g_{\rm rad}(z) = f(z)\int_{-\infty}^\infty \de x\; \dfrac{\phi(x)}{\sqrt{c +
\int_0^z \de z'\;\varrho(z')\;\phi(x-u(z'))}}.
\end{equation}
Here, $z$ is a planar coordinate pointing away from the star, $f(z)$ a function
that controls the stationary wind velocity law (and absorbs the \citet[hereafter
CAK]{cak} line force parameter $k$), $u(z)$ is the wind velocity law normalised
to the thermal speed, $v_{\rm th}$ (assumed to be constant), $\varrho(z)$ is the
wind density stratification, $x$ is the normalised frequency displacement from line
center in a resonance line with Doppler line profile $\phi(x)$, and $0<c\ll 1$ is a
cutoff parameter that prevents the wind from becoming completely transparent;
the argument of the square-root in the denominator of Eq.\ (\ref{grad}) is the
optical depth in a line with normalised and constant mass absorption coefficient,
$\kappa=1$, and the square-root corresponds to the second CAK force parameter
$\alpha$ being set to $\alpha=1/2$  (with this value chosen for computational
convenience). The line force (\ref{grad}) results from an integration over a
power-law distribution in frequency and mass absorption coefficient,
$N(\nu,\kappa) \sim \nu^{-1} \kappa^{-3/2}$. There is no angle integration in
Eq.\ (\ref{grad}), thus, it is in a single-ray approximation.

We fix the function $f(z)$ so that the resulting wind velocity law is
linear, $v\sim z$. The reason for this is that the steepness of the
CAK velocity law near the sonic point causes a very swift
length-stretching of the perturbations. The initial growth of the LDI
structure cannot be followed in sufficient detail thus, and nonlinear
effects appear to occur quite suddenly. Non-linear effects include the
excitation of a large number of harmonic overtones of the
perturbation and the mutual collisions and coalescence of density
enhancements (shell-shell-collisions); for more, see \citet{felsam}. By
assuming $u(z)\sim z$, the wind structure grows in arithmetic (velocity)
and geometric (density) progression as function of $z$ and all steps
of its evolution can be followed in detail.

This description of the line force shows the pure LDI without any
intervening or competing effects. Since we also neglect thermal gas
pressure and gravity, and since the radiative force in
Eq.\ (\ref{grad}) depends on the wind velocity and density alone, there is
thus a kinematical model for the wind and the LDI (if 
allowing for inertial mass in kinematics). Since the Sobolev
approximation has the velocity gradient entering $g_{\rm rad}$ in
Eq.\ (\ref{grad}) via $\sqrt{u'}$, the force in Eq.\ (\ref{grad})
still leads to an eigenvalue problem for the mass-loss rate in the
Euler equation with an X-type CAK critical point (not the sonic point)
that separates shallow ('breeze') solutions from steep ('overloaded'
or 'failed') ones.

The structure developing from the unstable growth of perturbations introduced at a
rather large amplitude at the inner boundary is at a maximum since scattering
is left out intentionally, thus there is no stabilising line drag effect
\citep{emlucy}. Furthermore, no Schuster-Schwarzschild absorbing layer is used
to reduce the radiative flux at small speeds on the order a few $v_{\text{th}}$
\citep[see again][]{ocr}. Finally, the sound speed and the gravity are set to
zero. Thus, there is  no barometric density stratification at the inner boundary
and the line force is the only force present (apart from numerical viscosity).
The effects of these simplifications are discussed in Sect.~\ref{disk}.

We performed hydrodynamic simulations using the code developed by
\citet{felto}. The code solves the Euler and continuity equations for a wind
subject to the line-driven instability from a pure absorption line force in
planar geometry. We use a simple time-explicit Euler scheme, where the density,
$\rho,$ and the momentum density, $\rho u,$ are derived from the continuity and
momentum equations solved in integral form. The fluxes at cell boundaries are
interpolated from their values inside the cells using the van Leer derivative
\citep{bram} and consistent advection \citep{norwi,zeus}.
The specific parameters and normalisations of our model are as
follows. We emphasise that given the simplicity of Eq.~(\ref{grad}),
the wind structure is generic, so any reference to a single, real star
would be putative and we can only refer to a 'generic' O supergiant
with a strong wind, for instance, $\zeta$~Pup or $\zeta$~Ori.

Velocities are given in units of $v_{\rm th}$. The latter enters the
(finite) thermal width of the line profile, which is the most
important factor for obtaining the LDI. Distance is normalised such that
the velocity law is not only linear, but $u(z)=z$. Mass units are
fixed by assuming a mass flux $\varrho u=1$ for the stationary
wind. With $z,u,$ and $m$, all units of our (extended) kinematics are
fixed (time $t=z/u$). As the inner boundary, we assume $z=1/3$ and $u=1/3$,
that is, one third of $v_{\rm th}$. Since the sound speed in hot-star
winds is typically a few $v_{\rm th}$, the inner boundary corresponds
to (roughly) 10\% the sound speed. The sonic point should be around
$z=u\approx 3$. We sample the wind structure in the following figures
at $z=1$ and $z=5$.

%Our hydrodynamics code is a simple time-explicit solver for the Euler and
%continuity equations: staggered meshes for $\varrho$ and $\varrho u$; control
%volume approach; van Leer flux interpolation at cell interfaces \citep{bram};
%consistent advection \citep{norwi,zeus}; numerical viscosity.
We use 1000 spatial
mesh points with $\Delta z/z=$~const. (logarithmic spacing). The Doppler profile
of the exemplaric spectral line is resolved by three frequency points per
thermal width and $2\times 3\times 3=18$ points in total.

As the opacity cutoff, we use $c=10^{-4}$, which is close to the value used
in most simulations thus far. The cutoff serves to avoid steep
acceleration of highly rarefied intershell gas. In \citet{felto},
we also used $c=10^{-10}$, showing this steep acceleration, but
with essentially no difference for the gas at the moderate densities that we are
interested here.

At the inner boundary, we assume perturbations in the mass flux
$\varrho u$ but keep the density $\varrho$ fixed. We implemented two
types of inner boundary perturbations, sinusoidal and Langevin
perturbations. The former can help us to understand the development of
wind structure and the latter provides a more realistic model for
atmospheric perturbations, which are likely to be the result of subsurface
convection and which contain many pulsational modes. We chose a
perturbation period and correlation period that are roughly equal to
the wind flow time and atmospheric cutoff period, which corresponds to the values given in
\citet{felpulpal}.

\subsection{Results of hydrodynamic simulations}

\begin{figure}
\includegraphics[width=0.5\textwidth]{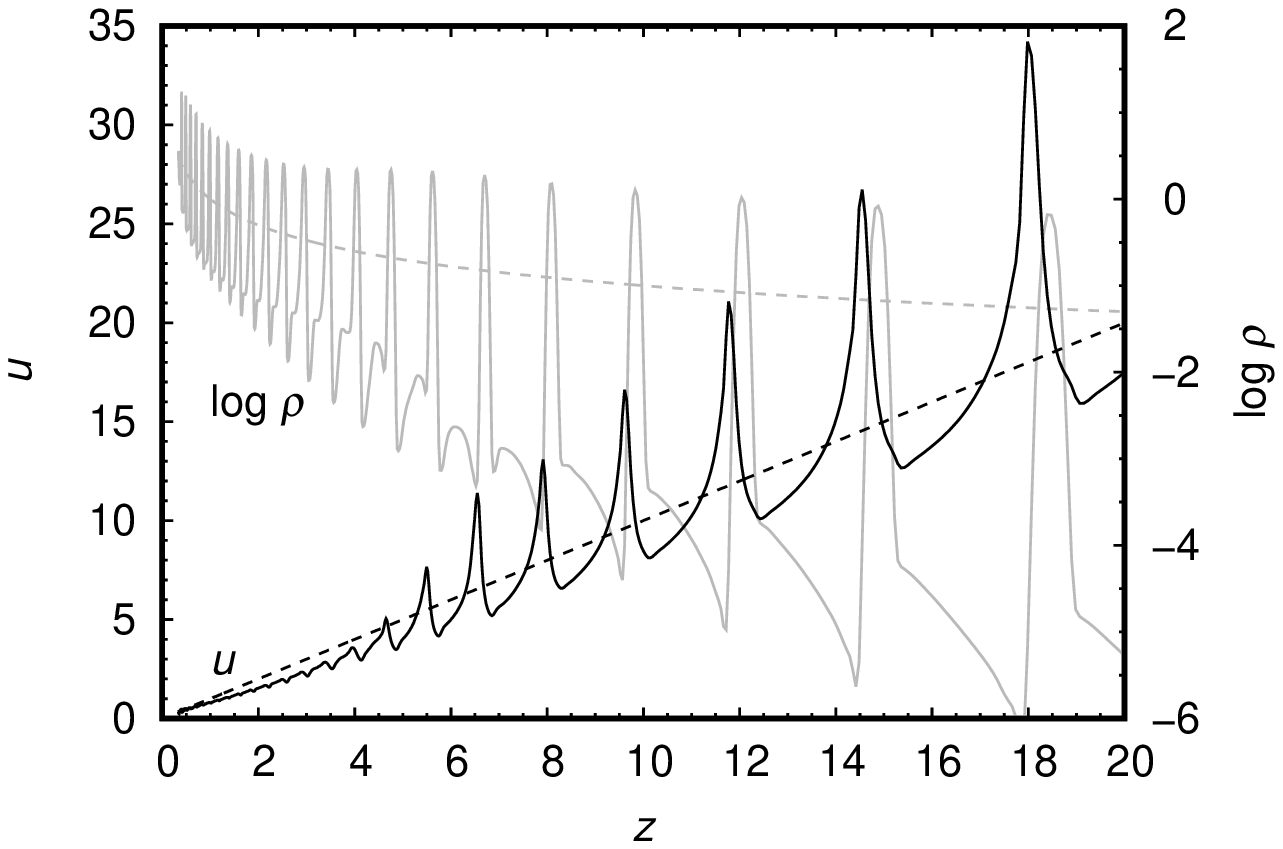}
\includegraphics[width=0.5\textwidth]{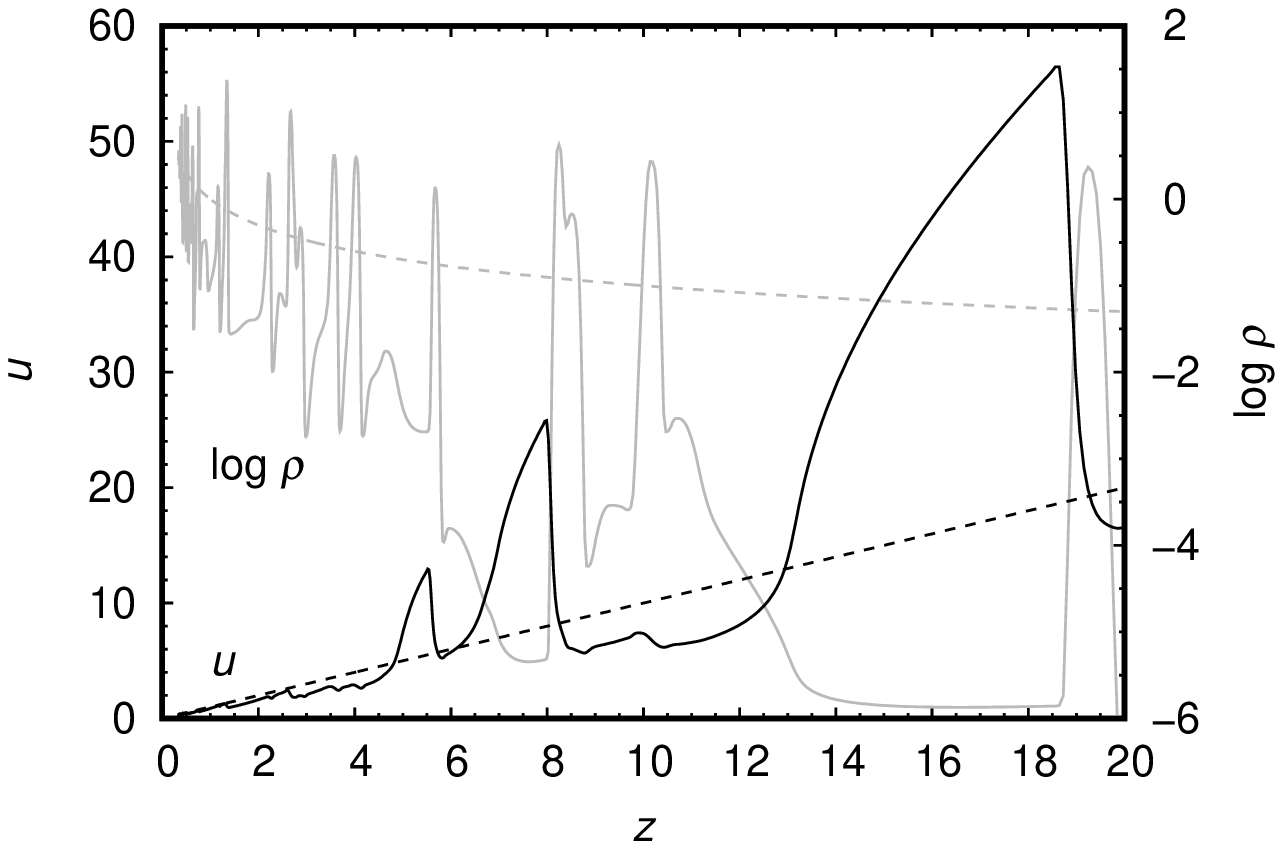}
\caption{Snapshot of height variations of the wind velocity (black lines) and
density (gray lines) compared with the stationary CAK solution (dashed lines), plotted for sinusoidal base perturbation ({\em upper panel}) and for the
Langevin base perturbation ({\em lower panel}). Calculations are
performed in non-dimensional units with velocity expressed in units of the
thermal speed.}
\label{radialsnap}
\end{figure}

Figure~\ref{radialsnap} shows snapshots of hydrodynamic simulations for
sinusoidal and Langevin base perturbations. The instability fully develops at a
velocity of several thermal speeds, leading to strong overdensities moving
roughly at the speed of the unperturbed wind. The rare\-fied matter between
dense clumps is strongly accelerated and accumulates in the closest overdensity
at larger height. Compared to sinusoidal base perturbations, the Langevin base
perturbation leads to a less regular structure of overdensities.

\begin{figure}
\includegraphics[width=0.5\textwidth]{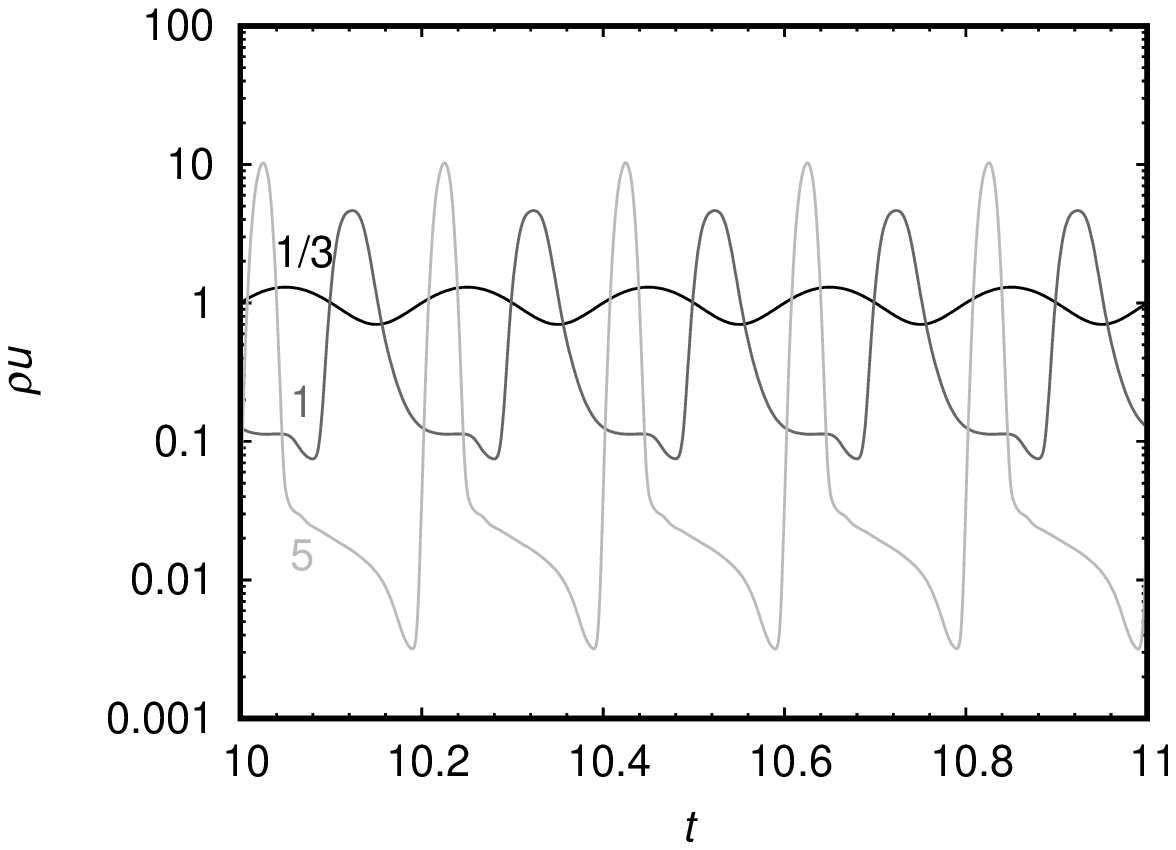}
\includegraphics[width=0.5\textwidth]{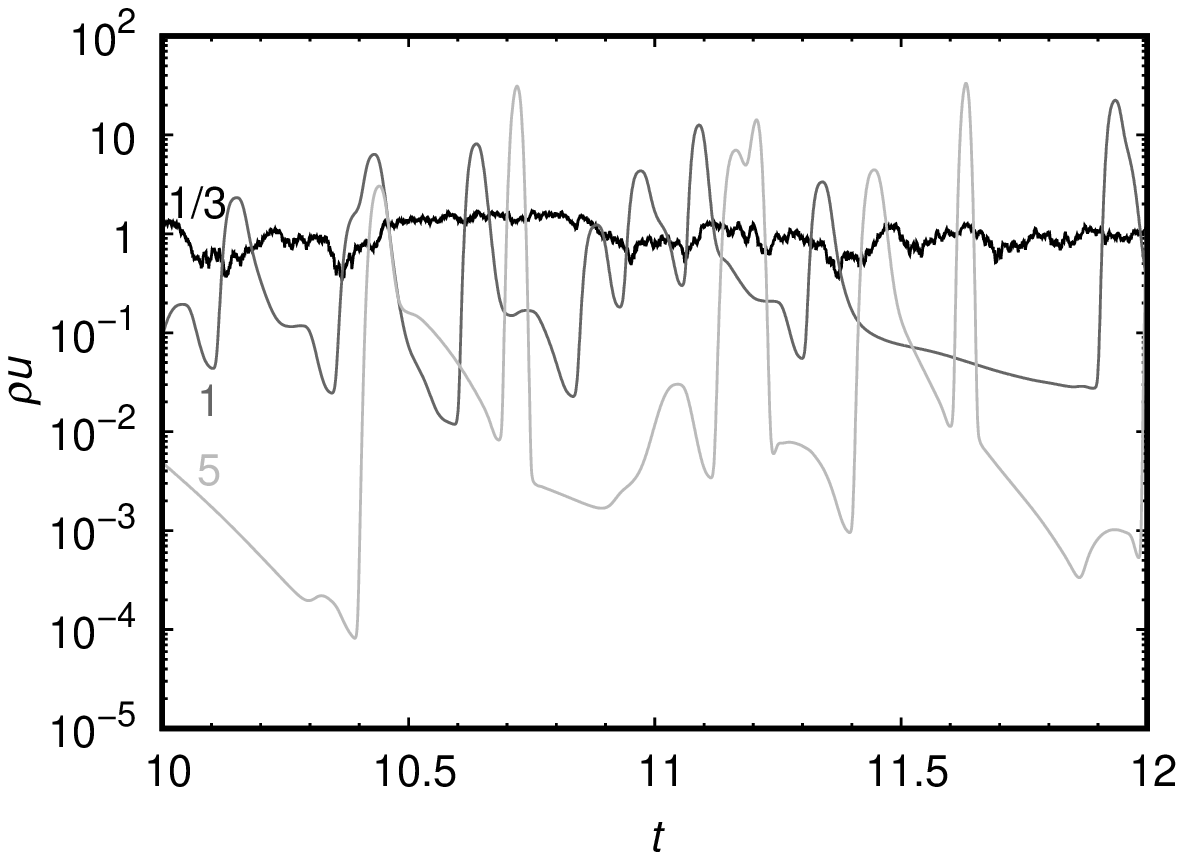}
\caption{Variability of the mass flux at $u=1/3$, $u=1$, and $u=5$ 
(increasing amplitude) for sinusoidal ({\em
upper panel}) and Langevin base perturbations ({\em lower panel}). 
Calculations are
performed in nondimensional units with stationary mass flux $\rho u=1$.}
\label{soundlanwave}
\end{figure}

We studied wind variations close to the star where the light variability
originates. For small base perturbations, the instability fully develops only at
very large velocities and, consequently, it does not cause large mass-loss rate
variations close to the photosphere. On the other hand, for large base
perturbations, the line-driven wind significantly affects the wind structure
already at velocities close to the thermal speed (see Fig.~\ref{soundlanwave}).
The amplitude of the boundary perturbation should be on the order of 0.3 -- 0.5
thermal speeds to cause significant variability at a low height.

The model with sinusoidal base perturbations (Fig.~\ref{soundlanwave}, upper
panel) shows the connection between boundary conditions and the development of
wind structure. Close to the boundary, the variations of the mass flux still
resemble the boundary perturbations. As the line-driven instability develops,
the maximum perturbations steepen into shocks collecting inner low-density
material moving at high speed.

This also provides the basic explanation of the structure shown for Langevin
boundary perturbations in Fig.~\ref{soundlanwave} (bottom panel). At high
velocities, only the largest overdensities prevail, originating from the
largest surface perturbations and engulfed subsequent smaller perturbations. 

\section{Modelling the light variability}

\subsection{Integrating over the stellar disk: Single surface patch}

\begin{figure}
\includegraphics[width=0.5\textwidth]{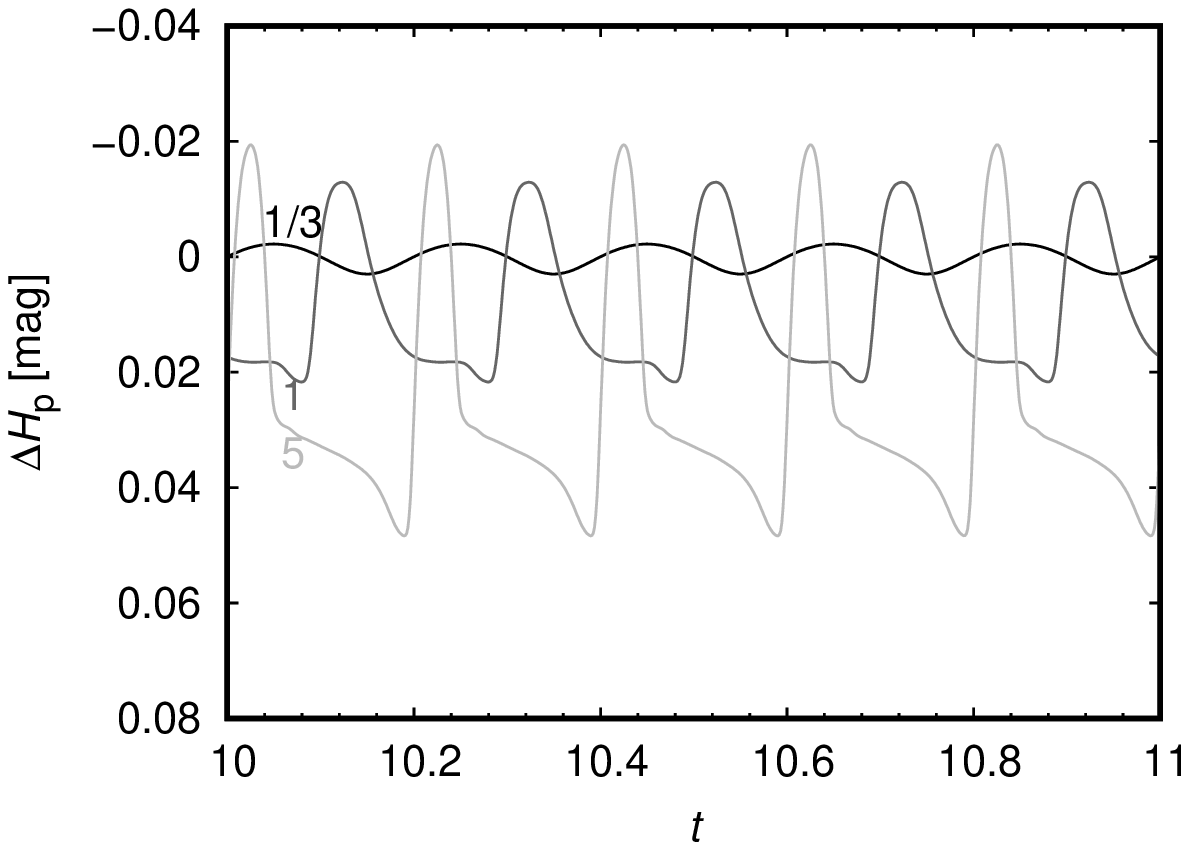}
\includegraphics[width=0.5\textwidth]{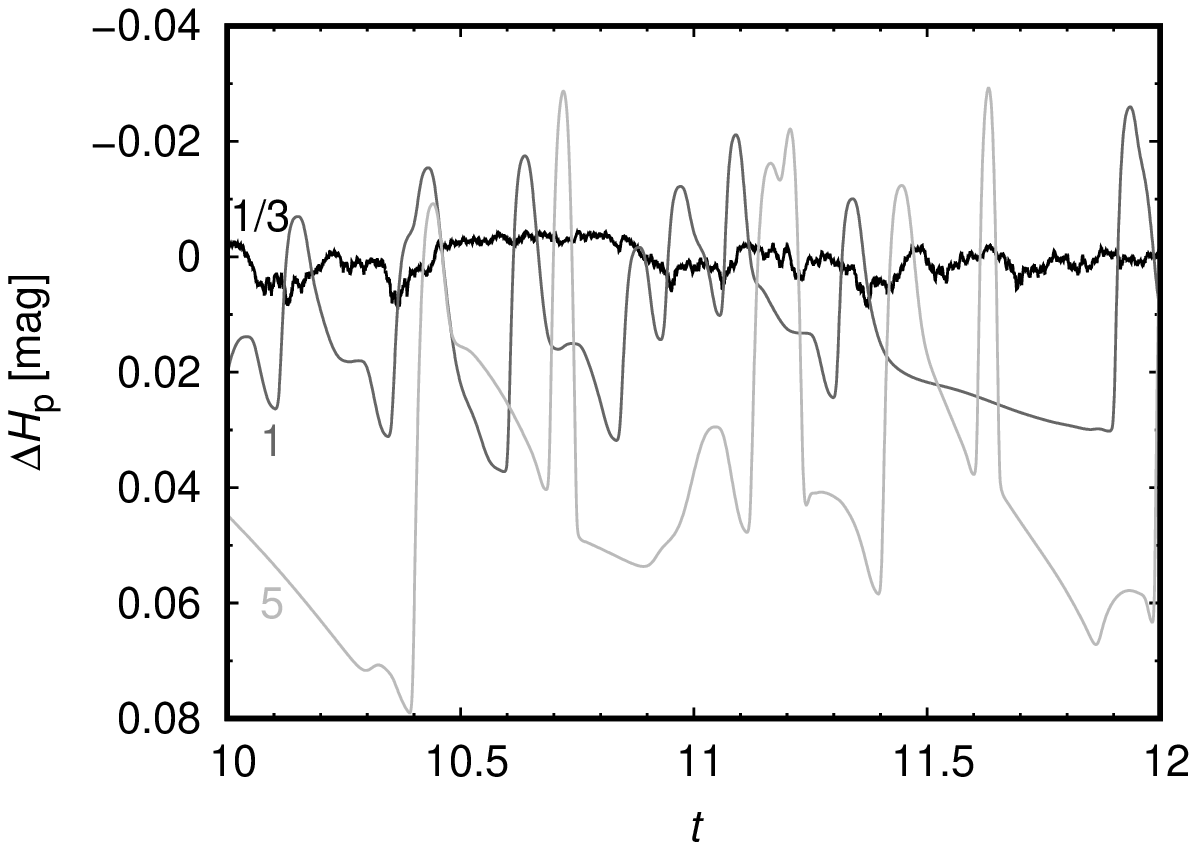}
\caption{Light variability for sound wave ({\em upper
panel}) and Langevin boundary perturbations ({\em bottom panel}).
Individual curves are calculated from the mass flux variations for
velocities $u=1/3$, $u=1$, and $u=5$. The graph shows the difference between
actual magnitude and magnitude predicted for the mean flux.}
\label{varn1}
\end{figure}

The predicted time variability of the mass flux is used to derive light
variations and light curves. We used the formula from \citet{magvar}, based on
the global spherically symmetric METUJE wind models and described by
\citet{cmfkont}, which connects the mass flux variations $\rho u$ at a
given height with photometric variability:
\begin{equation}
\label{hpt}
\Delta H_\text{p}=-2.5\log(e) \frac{\Delta F}{F_0}
\log\zav{\frac{\rho u}{\langle\rho u\rangle}}.
\end{equation}
Here, the fit parameter, $\Delta
F=1.6\times10^7\,\text{erg}\,\text{s}^{-1}\,\text{cm}^{-2}$, describes the
strength of the wind blanketing effect;
$F_0=9.0\times10^8\,\text{erg}\,\text{s}^{-1}\,\text{cm}^{-2}$ is the emergent
flux for a reference mass-loss rate; and $\langle\rho u\rangle$ is the mean mass
flux. This relation was derived for a typical O star with parameters
corresponding to HD~191612.

\begin{figure}
\includegraphics[width=0.5\textwidth]{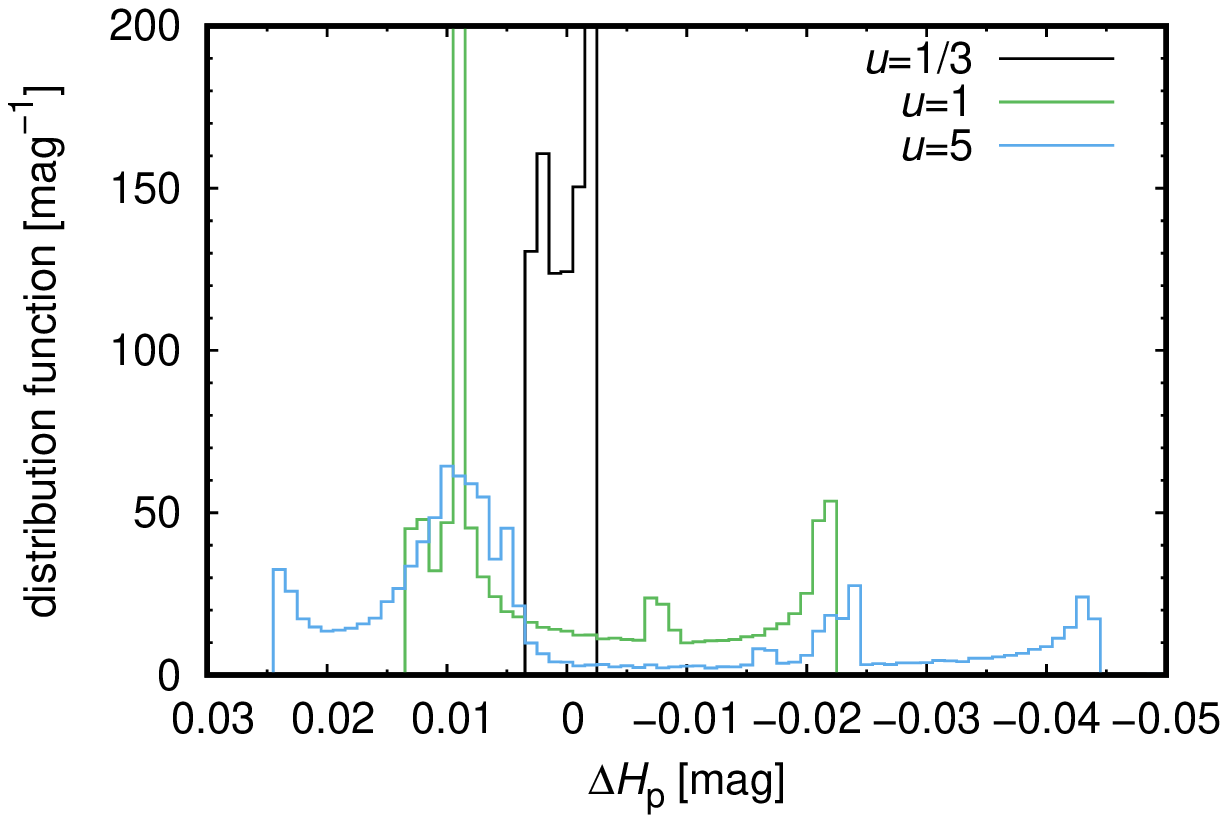}
\includegraphics[width=0.5\textwidth]{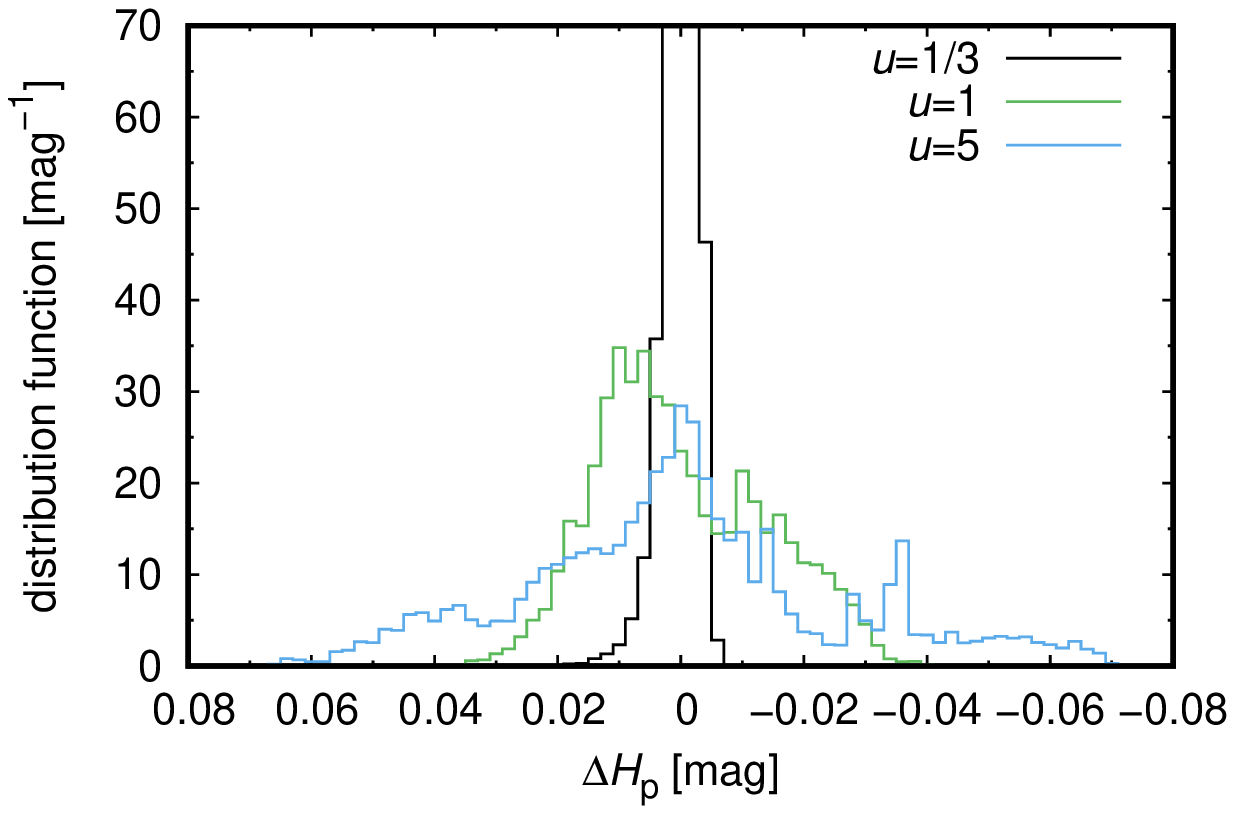}
\caption{Histograms of the distribution of magnitude differences for sound wave
perturbations ({\em upper panel}) and Langevin perturbations ({\em bottom
panel}), for $u=1/3$, $1$, and $5$. The ordinate gives the relative
frequency divided by the box width, i.e, the distribution function. It is
normalised in such a way that its integral over all magnitude differences is
unity.}
\label{hist}
\end{figure}

We calculated the light curves for both types of base perturbations (sinusoidal
and Langevin turbulence) and for mass flux variations at different heights (see
Fig.~\ref{varn1}). These variations closely follow the mass flux variations in
Fig.~\ref{soundlanwave}. For large boundary perturbations, the instability
develops sharp density peaks already in the subsonic part of the wind. These
peaks  also appear in the mass-loss rate variations and, consequently, also in
photometric variations due to wind blanketing. Because the peaks of the density
variations are not fully symmetric, the derived light variation becomes also
slightly asymmetric. Predicted sharp brightness peaks are typical for the light
variability of O stars connected with the line-driven wind instability and can,
therefore, provide an observational test of the appearance of the instability.

%\begin{figure}
%\includegraphics[width=0.5\textwidth]{langevin4r3.eps}
%\caption{Time-frequency diagram of the light variability for the case of the 
%Langevin base perturbation predicted at height $z=5$. The frequency spectrum was
%calculated beween a given time $t$ and $t+5$.}
%\label{langevin4r3}
%\end{figure}

\begin{figure}
\includegraphics[width=0.5\textwidth]{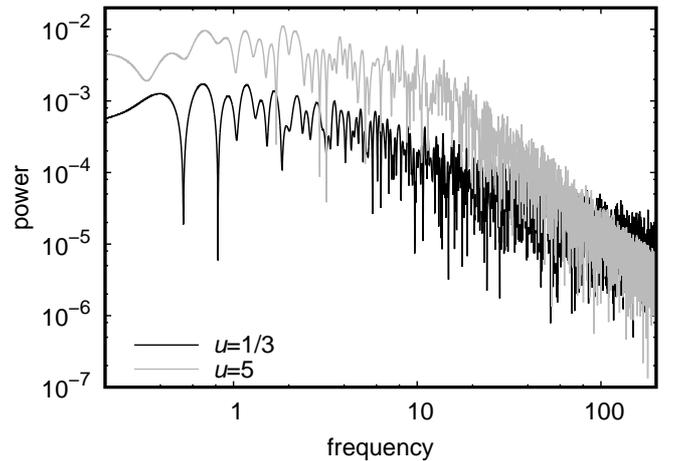}
\caption{Power spectrum of light variability for Langevin base perturbations,
for velocities $u=1/3$ and $u=5$. Determined as a magnitude of the coefficients
of the Fourier transform of the light curves.}
\label{langevin4er}
\end{figure}

Another signature of the line-driven wind instability is seen in the
distribution of magnitude differences in Fig.~\ref{hist}. The distribution at
the base of the wind given by the boundary conditions is significantly modified
by the line-driven instability. Not only does the distribution become wider,
but  its shape is modified as well. From Fig.~\ref{hist}, it follows that the peak
shifts to positive magnitude differences, while the tail formed by the
brightness maxima is more extended. This originates due to the structure
generated by the line-driven wind instability with sharp overdensity regions
where the sharp brightness maxima form, followed by broad underdensity regions
where the broad photometric minima have their origin.

This is also seen in the value of the skewness introduced as
$\gamma_1=\overline{(x-\overline{x})^3}/s^3$, where $x$ is variable, the
overline denotes mean value, and $s$ is the sample standard deviation. The
skewness becomes negative for large velocities ($\gamma_1=-0.27$ at $u=5$
for Langevin perturbations). Negative skewness implies that the magnitude
distribution has a tail on its brighter side, which is formed during the short period when
the overdensities appear, while most of the values appear on the dimmer side of
the distribution originating in the thin gas.

From the time-frequency analysis, it follows that the frequency spectrum is more
or less constant with time. On the other hand, the perturbations evolve as they
move in the wind. The radial development of instability structure can be
seen in the power spectrum of light variations in Fig.~\ref{langevin4er}.
While the perturbations close to the base are given by a uniform power law
due to the Langevin perturbations, the variability at large heights
shows a more complex, broken power law. As the high-frequency perturbations with
low density move faster, they collide and merge with less frequent high-density
perturbation. Consequently, the line-driving filters out power at higher
frequencies and maintains large low-frequency perturbations.

\subsection{Integrating over the stellar disk: Multiple surface patches}

So far, we have assumed that the stellar surface consists of a single patch and
that the wind structure does not depend on the lateral coordinates. This is a
very artificial assumption and we would expect that the boundary perturbations
triggering the line-driven instability are not coherent across the stellar
surface.

To account for the lateral fragmentation of the wind structure, we follow the approach
of \citet{desow} and \citet{lidarikala}, dividing the stellar wind into $N$
concentric spherical cones and assuming that the wind structure in each of
these cones is independent from neighboring cones. The mass flux at the time, $t,$ in
each cone is derived by randomly shifting the simulated mass flux over time:
\begin{equation}
(\rho u)_i(t)=\frac{\Omega_i}{4\pi}(\rho u)(t+\Delta t_i),
\end{equation}
where the index $i$ counts the cones, $\Omega_i$ is the solid angle subtended by
a cone, and $\Delta t_i$ is a random variable. We assume that $\Delta t_i$
has a uniform distribution over the simulation time, $\Delta T$, and that the
mass flux varies periodically with the same period $\Delta T$.
We selected the cones in such a way that the solid angle $\Omega_i$ is roughly
the same for each one and integrate the flux from all cones 
across the visible surface (see \citealt{oblavar} for details).
For the integration, we adopt the limb darkening coefficients from \citet{okraaj}.

Hydrodynamic simulations predict a typical lateral scale of structure generated
by the line-driven instability of about $0.01\,R_\ast$ \citep{sundsim}. This
corresponds to a patch size for wind structure on the order of several degrees, as also
inferred from emission line profile variability by \citet{desow}. A small
lateral size for the wind structure is also consistent with the number of clumps in the
observable part of the wind, which is on the order of $10^3$--$10^5$, as
estimated from the observed level of polarimetric variability \citep{davo}, from
the strength of wind line profiles \citep{osporcar,clres2}, and from the level
of short-term X-ray variability \citep{nazog}. Therefore, we selected $N=300$
and $N=10^3$ for the following simulations.

\begin{figure}
\includegraphics[width=0.5\textwidth]{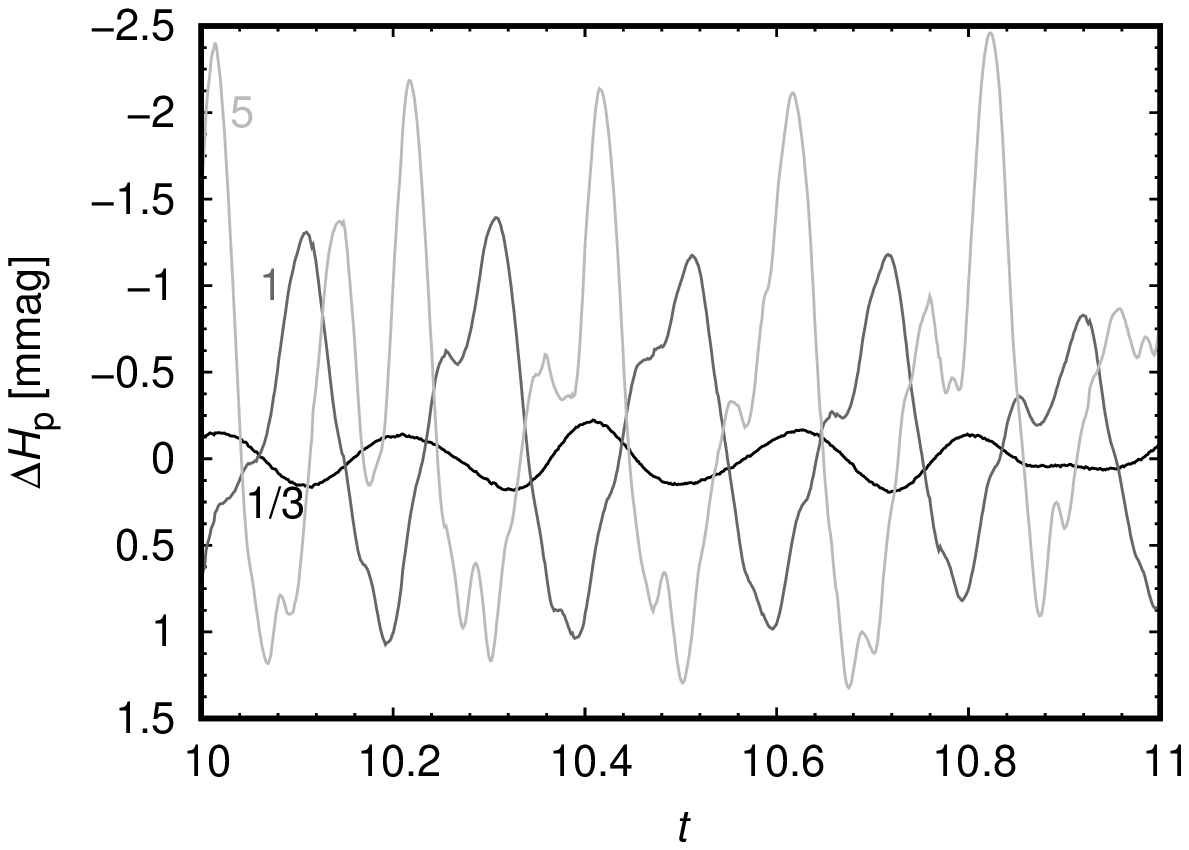}
\includegraphics[width=0.5\textwidth]{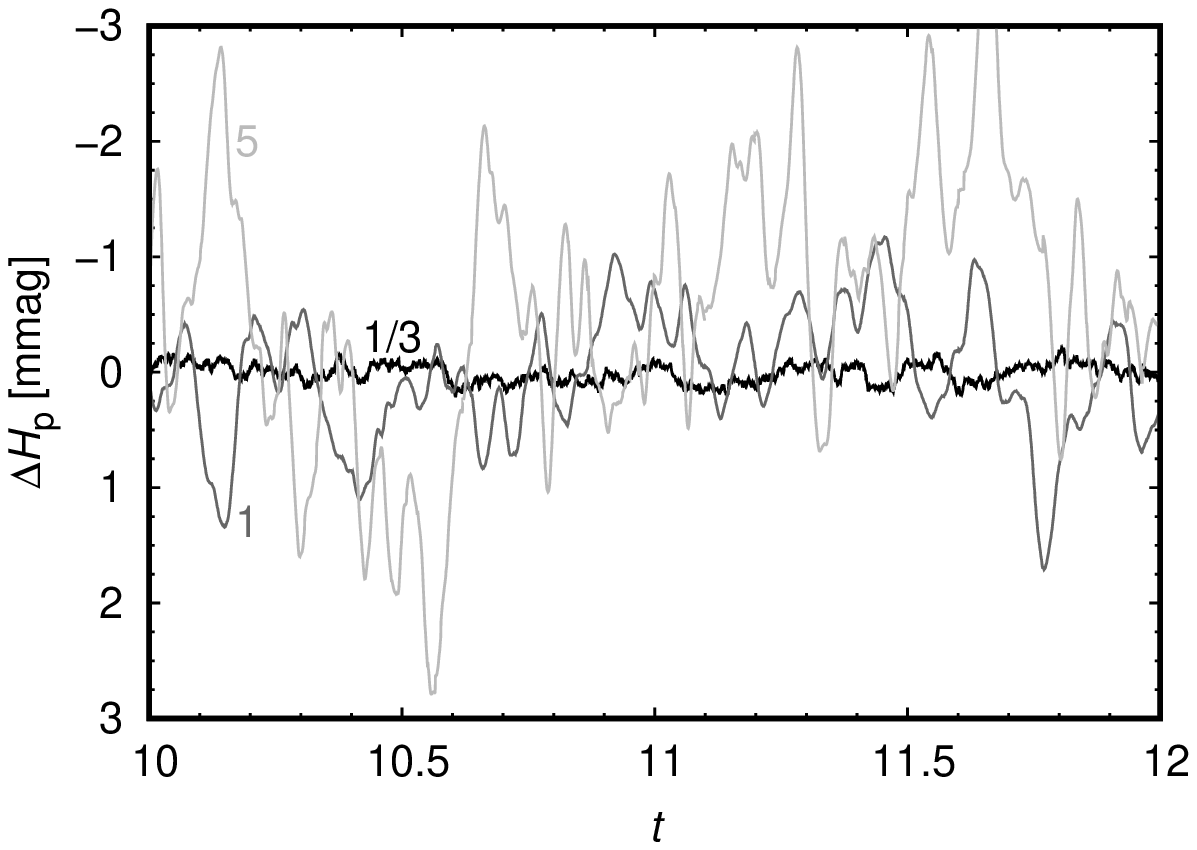}
\caption{Light variability for sound wave perturbations ({\em upper
panel}) and Langevin perturbations ({\em bottom panel}) summed over
$N=10^3$ independent cones, calculated from mass flux variations 
for $u=1/3$, $1$, and $5$. The graph shows the difference between 
actual magnitude and mean.}
\label{varn3}
\end{figure}

Obviously, as a result of summing over independent light curves, the amplitude
of the variability decreases with increasing $N$ \citep{oblavar}. However, the
variability remains detectable even for a relatively large number of cones,
$N=10^3$ (Fig.~\ref{varn3}). The form of the boundary perturbation has a significant
influence even on the combined light curves as the light curve for sound wave
boundary perturbations retains some kind of quasi-periodicity.

\begin{figure}
\includegraphics[width=0.5\textwidth]{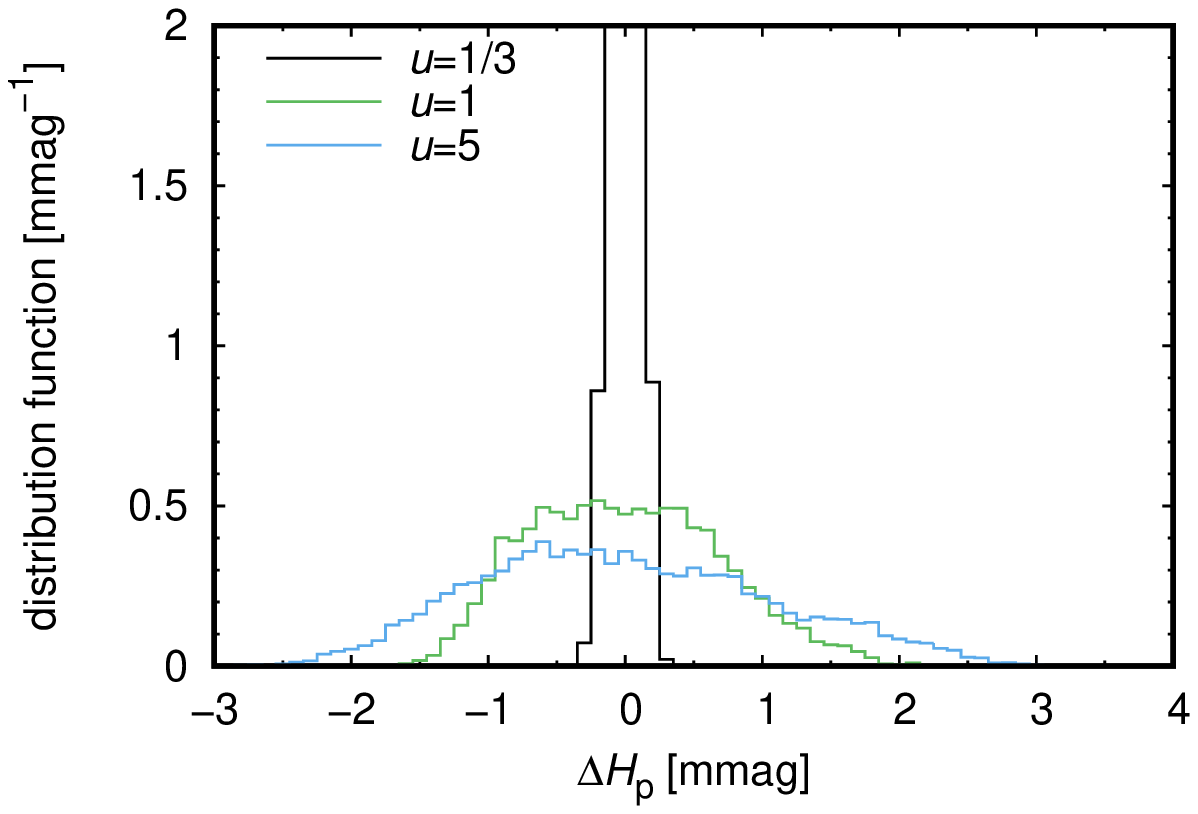}
\includegraphics[width=0.5\textwidth]{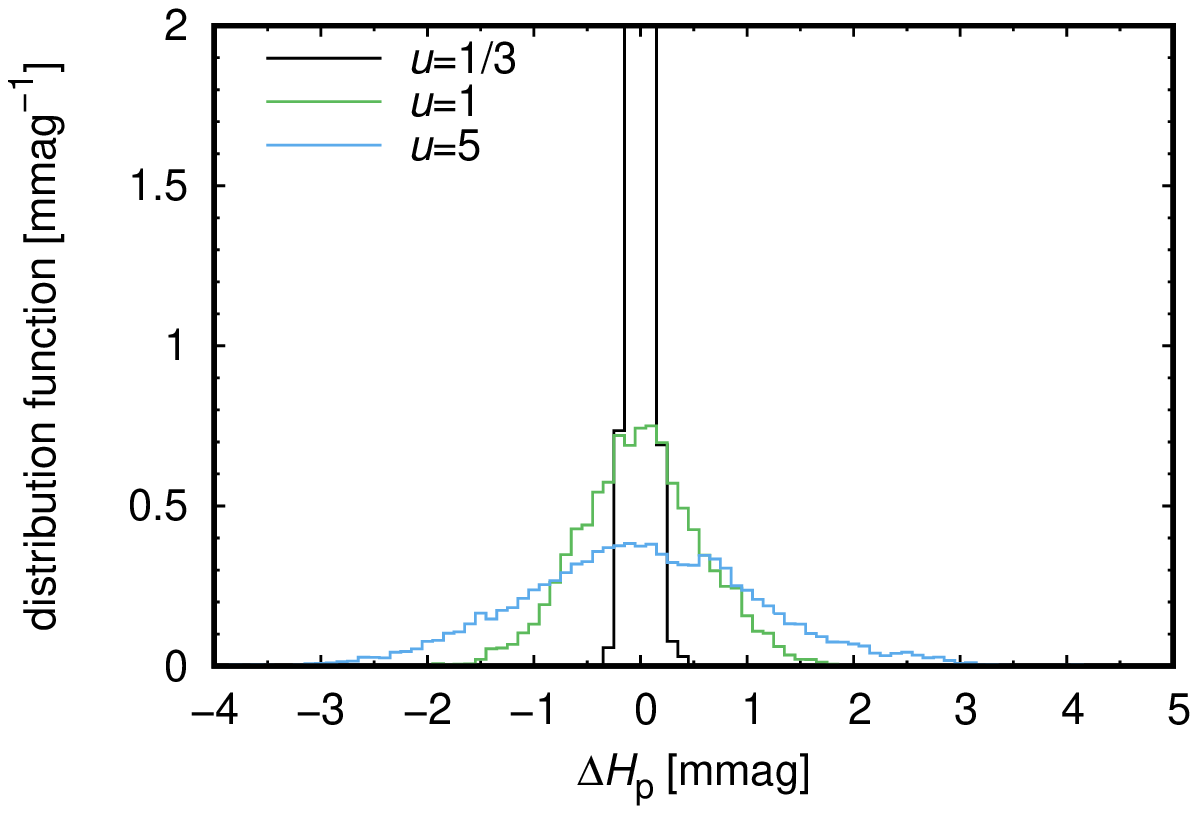}
\caption{Histogram of the distribution of magnitude differences for sound wave
perturbations ({\em upper panel}) and Langevin perturbations
({\em bottom panel}) summed over $N=10^3$ independent cones, calculated from 
mass flux variations at $u=1/3$, $1$, and $5$.} 
\label{varn3hist}
\end{figure}

%\begin{table}[t]
%\caption{Skewness $\gamma_1$ of the magnitude difference distribution for
%different base perturbations at different locations in the wind}
%\label{skewsimtab}
%\centering
%\begin{tabular}{ccc}
%\hline
%\hline
%Location & Sinusoidal & Langevin \\
%\hline
%\multicolumn{3}{c}{$N=250$}\\
%\hline
%$u=1/3$ & $-0.02$ & $0.07$  \\
%$u=1$   &  $0.28$ & $-0.02$ \\
%$u=5$   &  $0.28$ & $-0.06$ \\
%\hline
%\multicolumn{3}{c}{$N=1000$}\\
%\hline
%$u=1/3$ & $-0.06$ & $0.17$ \\
%$u=1$   &  $0.25$ & $0.04$ \\
%$u=5$   &  $0.25$ & $0.08$ \\
%\hline
%\end{tabular}
%\end{table}

After summing up over the independent cones, the distribution of magnitude
differences approaches a normal distribution (Fig.~\ref{varn3hist}). Still, even
for a large number of patches on the order of hundreds, the distribution resulting from the
Langevin base perturbation shows negative skewness. This indicates that the
magnitude distribution function has an extended tail on its brighter side. We
derived $\gamma_1=-0.030$ for $N=300$ and $\gamma_1=-0.015$ for $N=10^3$ at
$u=5$ and by averaging over $10^3$ random realisations of light curves. This
means that periods of lower flux (positive magnitude difference) formed in
underdensities are more frequent than periods of higher flux (negative magnitude
difference) formed in overdensities.
%At a wind base, for $u=1/3$, the skewness is given by the boundary
%perturbations, but it becomes positive in the wind. This means that the
%distribution of magnitude differences has tail for positive differences, while
%the star remains brigher more frequently. This is surprising, because the
%skewness due to spherically symmetric perturbations was negative. As a result
%of more regular pattern of variations in the case of soundwave boundary
%perturbation, the skewness is larger in this case.

\begin{figure}
\includegraphics[width=0.5\textwidth]{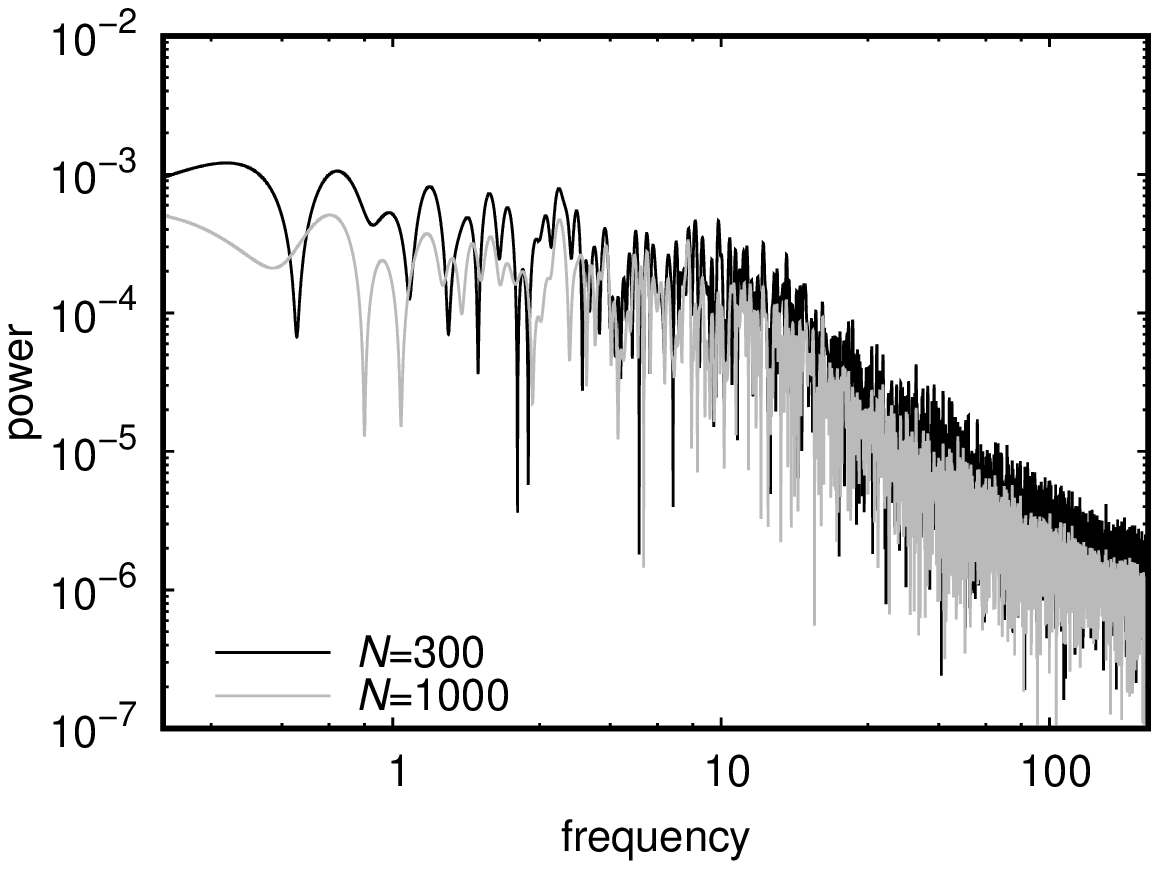}
\caption{Power spectrum derived from the light curve
summed over $N=300$ and $N=10^3$ independent cones (given in Fig.~\ref{varn3})
for the case of Langevin base perturbation at velocity $u=5$.}
\label{langevinvarn4er}
\end{figure}

The combination of light curves from individual patches also influences the
power spectrum of light variations significantly. The power spectrum of the
light curve derived from a single patch model shows a characteristic 'knee' as a
result of the merging of weak small-scale perturbations with large-scale
perturbations (Fig.~\ref{langevin4er}). This also occurs when the light
curve is combined from several independent cones (Fig.~\ref{langevinvarn4er}).

Finally, we briefly discuss the dependence of light variability on stellar
parameters. The height of continuum formation depends on the mass-loss rate.
With a rising mass-loss rate, the continuum formation region moves to the upper
layers of the atmosphere. Because the instability generated structure
strengthens with increasing height, we expect stronger stochastic variability in
stars with higher mass-loss rates.

\section{Testing against the TESS data}

%\begin{figure*}
%\includegraphics[width=\textwidth]{oblavar6}
%\caption{Sample of TESS light curves of studied OB stars.}
%\label{oblavar6}
%\end{figure*}

We tested the basic predictions of our model against the Transiting Exoplanet Survey
Satellite (TESS) data \citep{commander}. Using the MAST\footnote{Mikulski
Archive for Space Telescopes, https://archive.stsci.edu.} web page, we searched
for TESS photometric data for O stars contained in the Galactic O-Star Catalog
\citep{gosc}. We supplemented this list with B supergiants given in
\citet{crow}, \citet{lefever}, \citet{benagek}, \citet{markopulos}, and
\citet{hauci}. The search performed on September 17th 2020 resulted in 216
cross-matched light curves of O stars and 70 of B~supergiants, for which we
downloaded the light curves observed at a two-minute cadence. The visual
inspection of the data showed that all selected stars show some kind of light
variability, which is either periodic (that is, mostly due to binary effects or
pulsations), or stochastic, or a combination of both \citep{dalsipedersen,burs}.
We selected only stars with purely stochastic variability. We discarded stars
having periodical structures in their light curves, which can be attributed to
binarity and pulsations, as well as stars whose timescale of variability is comparable
to the duration of TESS observations. These effects especially appear in B
supergiants, which have larger radii than O stars with the same luminosity. This
reduced the list to 116 light curves of O stars and 18 of B~supergiants, for
which we performed a detailed analysis.

\begin{figure}
\includegraphics[width=0.5\textwidth]{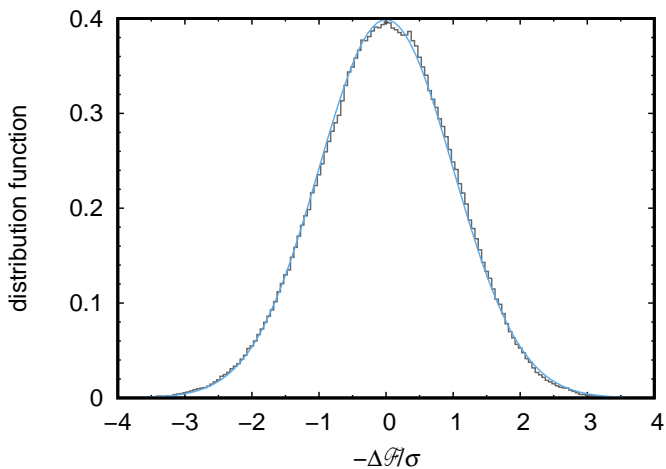}
\caption{Distribution of flux variations of TESS observations $\Delta {\cal F}$
relative to the dispersion $\sigma$ of individual light curves. The
distribution function is normalised to unity. The abscissa gives $-\Delta
{\cal F}/\sigma$, which is directly proportional to the magnitude difference.
The blue line denotes a Gaussian distribution.}
\label{oblavarhistall}
\end{figure}

Detailed analyses of the light curves show that the skewness of the
distribution function of magnitude differences is non-zero in most stars and it
appears to be independent of the stellar parameters. The light curves of most stars
have a negative skewness, indicating that dimmer phases prevail slightly  and that
the distribution has an extended tail on its bright side. A significant fraction
of stars show positive skewness and, therefore, the opposite behavior, but the
TESS light curves are rather short and the positive skewness may be just a
result of random selection of part of the light curve.
%This is obvious for stars for which the light curves from multiple visits are
%available. Such stars show large variations of skewness meaning that the
%scatter in Fig.~\ref{skewpar} is mostly the result of random variations.
%Therefore, we
We plotted the histogram of all magnitude differences relative to dispersion
(Fig.~\ref{oblavarhistall}). The final distribution has a negative skewness of
$\gamma_1=-0.084 \pm 0.002,$ corresponding to the mean skewness of all light
curves. This agrees with the light variations due to the line-driven wind
instability and a wind blanketing that is triggered by Langevin base
perturbations averaged over hundreds of independent surface patches.

All simulations of the line-driven instability so far show a sequence of highly
overdense and narrow shells or clumps, formed by the accumulation of gas from
broad and, as a result, highly rarefied regions between the shells or clumps. This
asymmetric density distribution reflects itself in an asymmetric distribution of
the visible stellar brightness via blanketing.

Therefore, the detected skewness of the stochastic light curves of OB stars offer direct evidence of a structure generated by line-driven wind instability that
already appears  close to the photosphere. Within our model, the negative
skewness of the light curves is caused by short periods when the light curve is
dominated by instability generated large overdensities that lead to brightening
of the star in the optical region (negative magnitude difference). These brief
periods of maximum light are followed by long periods of light minimum due to
rarefied wind regions that follow each overdensity.

%\begin{figure}
%\includegraphics[width=0.5\textwidth]{gamvsedt}
%\includegraphics[width=0.5\textwidth]{gamvsedf}
%\caption{Dependence of duration of individual peaks of light curves
%({\em upper panel}) and their skewness on the peak amplitude ({\em lower
%panel}).}
%\label{gamvse}
%\end{figure}

%We further analyzed peaks in light curves of each star separately. We used our
%own code to detect local maxima and minima of the light curves. The region of
%each peaks was selected as an interval where the observed flux is higher than
%the flux correspoding to the larger of the two neibhouring minima. We determined
%the amplitude of each peak $\Delta F$ as a difference between maximum and
%minimum, the duration of each peak $\Delta t$, and the skewness $\gamma_1$ of
%the peak. In total we detected about 19\,000 peaks, whose properties are plotted
%in Fig.~\ref{gamvse}. For individual light curves, there is linear correlation
%between the duration of each peak and its amplitude, which is apparent also for
%the whole sample. This indicates that all peaks are similar to some extent. The
%individual peak show on average positive skewness that seem to decrease with
%increasing amplitude. This could be interpreted as a result of combination of
%individual peaks from large sample of incoherent surface structures. Peaks with
%lower amplitude are formed from smaller number of surface patches on average,
%what enables them to keep their original skewness. On the other hand, larger
%peaks are formed from larger number of peaks on average, which lose their
%skewness.

\begin{figure}
\includegraphics[width=0.5\textwidth]{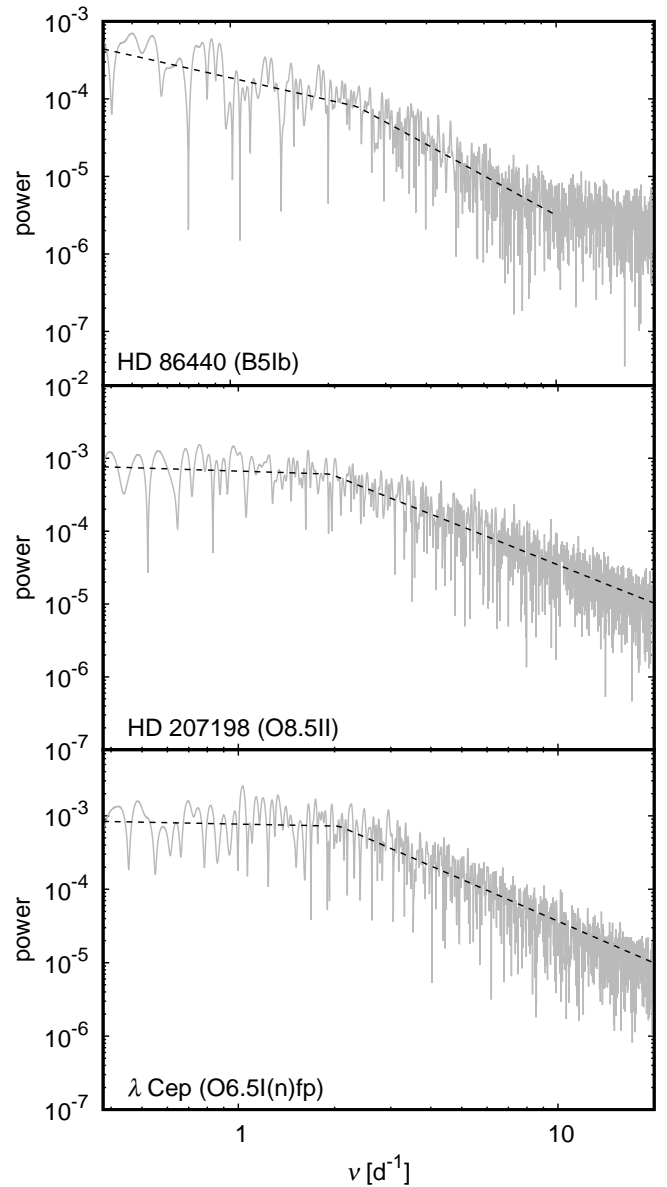}
\caption{Power spectrum of the observed TESS variations of three selected stars
(gray lines). Dashed black lines denote a fit using a broken power law.}
\label{twosfour}
\end{figure}

The power spectra of light curves of a significant fraction of stars are broken
power laws (see Fig.~\ref{twosfour}). Such power laws with a knee are expected
from our simulations as a result of the radial evolution of base perturbations,
during which small-scale perturbations are engulfed by large-scale
perturbations. Consequently, this feature can be also attributed to the
line-driven wind instability.

\section{Discussion}
\label{disk}

An advantage of the reduced description of the line force and wind
hydrodynamics introduced in Sect.~\ref{hydroeq} is that it gives the
'pure' LDI and thus a clear picture
of the generic structure that develops from the maximum possible
radiation force (no backscattering). This structure is highly
non-linear, with shock fronts of Mach number $\le 30$ (it is
questionable, however, how meaningful a reference of radiative shock
strength to thermal gas pressure realy is). Still, gas pressure and
radiative scattering, which is neglected in the present work, can have severe
influence on the wind dynamics. Here, we discuss a few of these effects.

Firstly, the so-called line-drag effect of \citet{emlucy} causes a partial
compensation of the LDI by backscattered photons. In the extreme case
that half the photons are backscattered, as may happen near the star,
the LDI has completely vanished. When using the SSF force in our
simulations (not shown in the present paper) with a backscattering
fraction of 1/2 throughout the wind, we obtain a wind that is
completely stable, base perturbations are not amplified, and the wind
stays near its stationary solution; there is thus no LDI structure at
all. Therefore, any quantitative estimate for the strength of wind
structure must include line scattering.

Secondly, by including thermal pressure and gravity, the wind velocity grows
exponentially up to the sonic point, then switches to the CAK wind
velocity law (the latter having an infinite slope at $r=R_\ast$).
This transition between different velocity laws is accompanied by
large changes in the velocity gradient, $u'$, which leads to fore-aft
asymmetries in the escape probabilities (the larger $u'$, the lower is
the Sobolev optical depth in the line). The importance of this shows
up most prominently in the EISF formulation of the radiative force,
with a source function $S=\beta_c/\beta$, where $\beta$ is the escape
probability in any direction, $\beta_c$ is the same but for directions towards
the stellar disk, and where $\beta$ and $\beta_c$ are calculated from
the actual, time-dependent wind structure, thus allowing for another
feedback (apart from that of the optical depth on the line force as in
Eq.\ \ref{grad}) of the hydrodynamics on the radiation field. \citet{owpu}
find that a depression in the source function near the
sonic point alters the mean wind conditions. The EISF models
settle to the unique steep solution of a nodal-type critical point.
\citet{sunowroz} find that with $S\rightarrow 0$, the
solution topology near the sonic point changes from X-type (as in CAK)
to nodal type, the latter having a steep branch through which a unique
solution passes and a shallow branch with a `degenerate' family of
solutions (all with the same $u'$).

Thirdly, this nodal solution topology at the sonic point was already
found by \citet{havran} for a pure absorption line
model. They find that at $v_{\rm th}/a=1/2$, the wind settles in the
unique steep branch at the node and the overall velocity law appears
to be too fast; whereas for realistic values, $v_{\rm th}/a=1/3$, the
wind favors the lower node branch that has a whole solution family.
Their degeneracy causes intrinsic, self-excited wind variability with
the wind sampling through all possible solutions.

Next, by including thermal pressure, perturbations grow exponentially
in the subsonic wind region already without any radiative force (by
pure energy conservation of a sound wave in a barometric
stratification),  eventually leading to so-called shock levitation in
dust-driven winds from cool stars. Sound waves then travel forth and
back through the wind between the inner boundary and the sonic
point. This has some poorly understood influence on the stability of
the inner wind. Current simulations have to adopt therefore a
photospheric speed of roughly 10\% the sound speed. Using 1\% or 30\%
instead, the simulations show huge oscillations and usually crash.

Therefore,  both the pure absorption model and the models including
scattering (SSF and EISF) show complications at the critical point,
with transitions from X-type to nodal type solution topology, and with
possible consequences for the overall wind dynamics (a stable steep
solution versus a\ degenerate, self-excited shallow solution). In contrast, our pure absorption models without thermal gas pressure that are
presented here show no peculiarities. Without explicit perturbations
at the wind base, the wind remains perfectly stable for the $u=z$
solution. With a periodic base perturbation, the unstable structure
from the line force in Eq.\ (\ref{grad}) gives a temporally
well-resolved and uniform growth of unstable structure from linear to
highly nonlinear phases.  Since we are interested in the velocity
law at small distances from the star near the sonic point, this
conditioning of the velocity law $u(z)$ to $u(z)=z$ should have no
significant influence on our results, but this makes the numerics much more
robust compared to calculations including thermal pressure and
scattering.

The resulting structure is the one that was first described by \citet{ocr}
and \citet{owosam}: the LDI enhances positive
perturbations of $u'$ and accelerates corresponding material until it
is eventually decelerated by the gas preceding it, which has either no or negative
velocity-gradient perturbations and thus remains at the stationary CAK
wind speed. The deceleration of gas happens in a strong reverse shock
with huge density contrasts (assuming there are no thermal energy
losses, i.e. there is no gas heating). We note in passing that
earlier simulations of the pure absorption line force encountered certain
difficulties caused by the very strength of the LDI, which were
overcome, for instance, by a Schuster-Schwarzschild absorbing layer in the
photosphere and a rather low opacity cutoff. The present models shows
no such problems.

A basic question of the present paper is for the maximum
perturbation amplitude the LDI can form near the sonic point. The
velocity difference caused by acceleration due to the instability is
on the order of the stationary wind speed. Thus, we may expect that
shocks with a velocity jump up to a few sound speeds may form
slightly above the sonic point. We find that with all intervening
effects from scattering (etc.)\ left out, such shocks can indeed occur.

We emphasise that a serious drawback of our simplifications is that no
meaningful inclusion of scattering seems possible. We already fixed the
velocity law of the wind arbitrarily to $u=z$ with the function
$f(z)$, and we would have to choose another arbitrary function, $0\le
g(z)\le 0.5$, for the fraction of back-scattering. We see no
unambiguous way to do this, especially in the highly significant
region close to the star. The results of the present paper have
therefore to be taken with caution since this pure absorption line
force will definitely overestimate the amount of structure formed by
the LDI in a more realistic setting.

We note, however, that such an approach is rather customary for
simulations of the LDI thus far, where  an optically thin source
function has typically been assumed (geometric dilution only). If, instead, a simple
approximation to an optically thick source function were used, the
unstable structure would be strongly reduced. For more, see \citet{felphd}.

\section{Summary}

We studied the photometric signatures resulting from the line-driven wind
instability by applying either sinusoidal or stochastic boundary perturbations.
We determined the photometric variability from mass-flux variations, assuming
that the wind consists of a large number of independent conical sectors.

The calculations show that the instability already develops  around the sonic
point of the wind, provided that the base perturbations are comparable to (but
smaller than) the thermal speed. Because the light variability originates
in the subsonic part of the wind, this means that the observed light variability
of O stars can be caused by wind variations if the base perturbations are
sufficiently large.

We find two signatures of the line-driven wind instability in our simulated
photometric data: a knee in the power spectrum of magnitude fluctuations and a
negative skewness of the distribution of magnitude fluctuations. The knee can be
explained by the engulfment of small-scale perturbations by large overdensities,
while the negative skewness is the result of spatial dominance of rarefied
regions, which leads to lower flux and higher magnitude difference. Both the
knee and the negative skewness endure when combining the light curves from
independent wind sectors, even for a relatively large number of sectors.

The simulated light curves are compared with TESS light curves of OB stars that
show stochastic variability. The observed variations bear signatures of the
line-driven wind instability. The distribution function of magnitude differences
of observational light curves shows negative skewness implying more frequent
periods of lower flux.
%Negative skewness can be explained as a result of the
%wind structure resulting from stochastic base perturbations.
Moreover, the power
spectrum of a significant fraction of light curves show a knee.
%, which can be
%attributed to the engulfment of small-scale perturbations by large-scale
%perturbations.
These observational features could be direct evidence for a
ensemble of discrete overdensities embedded in large rarefied regions created by
line-driven wind instability.

Although wind blanketing due to the line-driven wind instability provides a
reasonable explanation of stochastic light variability of hot stars, there may
be other competing processes leading to light variability, such as subsurface convection. However, additional observational tests of the
mechanism ought to be undertaken. Most of the light absorbed by the wind comes from the
extreme ultraviolet region (which is hardly accessible for observation). This
means that far-ultraviolet photometry should mostly vary in phase with visual
observations. Ultraviolet and visible variations in phase are not expected due
to hot convective plumes, which are expected to be manifested mostly as temperature
variations. Moreover, there are regions in the far ultraviolet domain where the
flux varies in antiphase with visual variability \citep[e.g. around
1300\,\AA,][]{magvar}. This provides an additional observational test of wind
blanketing variability.

\begin{acknowledgements}
The authors thank Prof.~Zden\v ek Mikul\'a\v sek for the discussion of
TESS photometric light curves.
This article is based upon work from the “ChETEC” COST Action (CA16117),
supported by COST (European Cooperation in Science and Technology). JK
acknowledges support by grant GA\,\v{C}R 18-05665S. 
\end{acknowledgements}

\bibliographystyle{aa}
\bibliography{papers}

\begin{thebibliography}{58}
\expandafter\ifx\csname natexlab\endcsname\relax\def\natexlab#1{#1}\fi

\bibitem[{{Abbott} \& {Hummer}(1985)}]{acko}
{Abbott}, D.~C. \& {Hummer}, D.~G. 1985, \apj, 294, 286

\bibitem[{{Aerts} {et~al.}(2009){Aerts}, {Puls}, {Godart}, \& {Dupret}}]{ae}
{Aerts}, C., {Puls}, J., {Godart}, M., \& {Dupret}, M.~A. 2009, \aap, 508, 409

\bibitem[{{Aerts} \& {Rogers}(2015)}]{aero}
{Aerts}, C. \& {Rogers}, T.~M. 2015, \apjl, 806, L33

\bibitem[{{Aerts} {et~al.}(2017){Aerts}, {S{\'\i}mon-D{\'\i}az}, {Bloemen},
  {Debosscher}, {P{\'a}pics}, {Bryson}, {Still}, {Moravveji}, {Williamson},
  {Grundahl}, {Fredslund Andersen}, {Antoci}, {Pall{\'e}},
  {Christensen-Dalsgaard}, \& {Rogers}}]{asid}
{Aerts}, C., {S{\'\i}mon-D{\'\i}az}, S., {Bloemen}, S., {et~al.} 2017, \aap,
  602, A32

\bibitem[{{Benaglia} {et~al.}(2007){Benaglia}, {Vink}, {Mart{\'\i}}, {Ma{\'\i}z
  Apell{\'a}niz}, {Koribalski}, \& {Crowther}}]{benagek}
{Benaglia}, P., {Vink}, J.~S., {Mart{\'\i}}, J., {et~al.} 2007, \aap, 467, 1265

\bibitem[{{Blomme} {et~al.}(2011){Blomme}, {Mahy}, {Catala}, {Cuypers},
  {Gosset}, {Godart}, {Montalban}, {Ventura}, {Rauw}, {Morel}, {Degroote},
  {Aerts}, {Noels}, {Michel}, {Baudin}, {Baglin}, {Auvergne}, \&
  {Samadi}}]{blomcor}
{Blomme}, R., {Mahy}, L., {Catala}, C., {et~al.} 2011, \aap, 533, A4

\bibitem[{{Bohannan} {et~al.}(1986){Bohannan}, {Abbott}, {Voels}, \&
  {Hummer}}]{boh}
{Bohannan}, B., {Abbott}, D.~C., {Voels}, S.~A., \& {Hummer}, D.~G. 1986, \apj,
  308, 728

\bibitem[{{Burssens} {et~al.}(2020){Burssens}, {Sim{\'o}n-D{\'\i}az}, {Bowman},
  {Holgado}, {Michielsen}, {de Burgos}, {Castro}, {Barb{\'a}}, \&
  {Aerts}}]{burs}
{Burssens}, S., {Sim{\'o}n-D{\'\i}az}, S., {Bowman}, D.~M., {et~al.} 2020,
  \aap, 639, A81

\bibitem[{{Cantiello} {et~al.}(2009){Cantiello}, {Langer}, {Brott}, {de Koter},
  {Shore}, {Vink}, {Voegler}, {Lennon}, \& {Yoon}}]{kant}
{Cantiello}, M., {Langer}, N., {Brott}, I., {et~al.} 2009, \aap, 499, 279

\bibitem[{{Carlberg}(1980)}]{karl}
{Carlberg}, R.~G. 1980, \apj, 241, 1131

\bibitem[{{Castor} {et~al.}(1975){Castor}, {Abbott}, \& {Klein}}]{cak}
{Castor}, J.~I., {Abbott}, D.~C., \& {Klein}, R.~I. 1975, \apj, 195, 157

\bibitem[{{Crowther} {et~al.}(2002){Crowther}, {Hillier}, {Evans}, {Fullerton},
  {De Marco}, \& {Willis}}]{prvnifosfor}
{Crowther}, P.~A., {Hillier}, D.~J., {Evans}, C.~J., {et~al.} 2002, \apj, 579,
  774

\bibitem[{{Crowther} {et~al.}(2006){Crowther}, {Lennon}, \& {Walborn}}]{crow}
{Crowther}, P.~A., {Lennon}, D.~J., \& {Walborn}, N.~R. 2006, \aap, 446, 279

\bibitem[{{David-Uraz} {et~al.}(2017){David-Uraz}, {Owocki}, {Wade},
  {Sundqvist}, \& {Kee}}]{duo}
{David-Uraz}, A., {Owocki}, S.~P., {Wade}, G.~A., {Sundqvist}, J.~O., \& {Kee},
  N.~D. 2017, \mnras, 470, 3672

\bibitem[{{Davies} {et~al.}(2007){Davies}, {Vink}, \& {Oudmaijer}}]{davo}
{Davies}, B., {Vink}, J.~S., \& {Oudmaijer}, R.~D. 2007, \aap, 469, 1045

\bibitem[{{Dessart} \& {Owocki}(2002)}]{desow}
{Dessart}, L. \& {Owocki}, S.~P. 2002, \aap, 383, 1113

\bibitem[{{Dufton} {et~al.}(2006){Dufton}, {Ryans}, {Sim{\'o}n-D{\'\i}az},
  {Trundle}, \& {Lennon}}]{dufbvele}
{Dufton}, P.~L., {Ryans}, R.~S.~I., {Sim{\'o}n-D{\'\i}az}, S., {Trundle}, C.,
  \& {Lennon}, D.~J. 2006, \aap, 451, 603

\bibitem[{{Feldmeier}(1993)}]{felphd}
{Feldmeier}, A. 1993, PhD thesis, Ludwig-Maximilians-Universit\"at, M\"unchen

\bibitem[{{Feldmeier}(1995)}]{felsam}
{Feldmeier}, A. 1995, \aap, 299, 523

\bibitem[{{Feldmeier} {et~al.}(2003){Feldmeier}, {Oskinova}, \&
  {Hamann}}]{lidarikala}
{Feldmeier}, A., {Oskinova}, L., \& {Hamann}, W.-R. 2003, \aap, 403, 217

\bibitem[{{Feldmeier} {et~al.}(1997){Feldmeier}, {Puls}, \&
  {Pauldrach}}]{felpulpal}
{Feldmeier}, A., {Puls}, J., \& {Pauldrach}, A.~W.~A. 1997, \aap, 322, 878

\bibitem[{{Feldmeier} \& {Thomas}(2017)}]{felto}
{Feldmeier}, A. \& {Thomas}, T. 2017, \mnras, 469, 3102

\bibitem[{{Haucke} {et~al.}(2018){Haucke}, {Cidale}, {Venero}, {Cur{\'e}},
  {Kraus}, {Kanaan}, \& {Arcos}}]{hauci}
{Haucke}, M., {Cidale}, L.~S., {Venero}, R.~O.~J., {et~al.} 2018, \aap, 614,
  A91

\bibitem[{{Jiang} {et~al.}(2015){Jiang}, {Cantiello}, {Bildsten}, {Quataert},
  \& {Blaes}}]{jian}
{Jiang}, Y.-F., {Cantiello}, M., {Bildsten}, L., {Quataert}, E., \& {Blaes}, O.
  2015, \apj, 813, 74

\bibitem[{{Kholtygin} {et~al.}(2011){Kholtygin}, {Sudnik}, {Burlakova}, \&
  {Valyavin}}]{khos}
{Kholtygin}, A.~F., {Sudnik}, N.~P., {Burlakova}, T.~E., \& {Valyavin}, G.~G.
  2011, Astronomy Reports, 55, 1105

\bibitem[{{Kourniotis} {et~al.}(2014){Kourniotis}, {Bonanos}, {Soszy{\'n}ski},
  {Poleski}, {Krikelis}, {Udalski}, {Szyma{\'n}ski}, {Kubiak},
  {Pietrzy{\'n}ski}, {Wyrzykowski}, {Ulaczyk}, {Koz{\l}owski}, \&
  {Pietrukowicz}}]{kourbos}
{Kourniotis}, M., {Bonanos}, A.~Z., {Soszy{\'n}ski}, I., {et~al.} 2014, \aap,
  562, A125

\bibitem[{{Krti{\v c}ka} \& {Kub{\'a}t}(2017)}]{cmfkont}
{Krti{\v c}ka}, J. \& {Kub{\'a}t}, J. 2017, \aap, 606, A31

\bibitem[{{Krti{\v{c}}ka}(2016)}]{magvar}
{Krti{\v{c}}ka}, J. 2016, \aap, 594, A75

\bibitem[{{Krti{\v{c}}ka} \& {Feldmeier}(2018)}]{oblavar}
{Krti{\v{c}}ka}, J. \& {Feldmeier}, A. 2018, \aap, 617, A121

\bibitem[{{Lecoanet} {et~al.}(2019){Lecoanet}, {Cantiello}, {Quataert},
  {Couston}, {Burns}, {Pope}, {Jermyn}, {Favier}, \& {Le Bars}}]{leco}
{Lecoanet}, D., {Cantiello}, M., {Quataert}, E., {et~al.} 2019, \apjl, 886, L15

\bibitem[{{Lefever} {et~al.}(2007){Lefever}, {Puls}, \& {Aerts}}]{lefever}
{Lefever}, K., {Puls}, J., \& {Aerts}, C. 2007, \aap, 463, 1093

\bibitem[{{Lucy}(1984)}]{emlucy}
{Lucy}, L.~B. 1984, \apj, 284, 351

\bibitem[{{Lucy} \& {Solomon}(1970)}]{lusol}
{Lucy}, L.~B. \& {Solomon}, P.~M. 1970, \apj, 159, 879

\bibitem[{{MacGregor} {et~al.}(1979){MacGregor}, {Hartmann}, \&
  {Raymond}}]{MacGregor}
{MacGregor}, K.~B., {Hartmann}, L., \& {Raymond}, J.~C. 1979, \apj, 231, 514

\bibitem[{{Ma{\'\i}z Apell{\'a}niz} {et~al.}(2013){Ma{\'\i}z Apell{\'a}niz},
  {Sota}, {Morrell}, {Barb{\'a}}, {Walborn}, {Alfaro}, {Gamen}, {Arias}, \&
  {Gallego Calvente}}]{gosc}
{Ma{\'\i}z Apell{\'a}niz}, J., {Sota}, A., {Morrell}, N.~I., {et~al.} 2013, in
  Massive Stars: From alpha to Omega, 198

\bibitem[{{Markova} \& {Puls}(2008)}]{markopulos}
{Markova}, N. \& {Puls}, J. 2008, \aap, 478, 823

\bibitem[{{Martins} {et~al.}(2015){Martins}, {Marcolino}, {Hillier}, {Donati},
  \& {Bouret}}]{marmarvar}
{Martins}, F., {Marcolino}, W., {Hillier}, D.~J., {Donati}, J.~F., \& {Bouret},
  J.~C. 2015, \aap, 574, A142

\bibitem[{{Martins} {et~al.}(2005){Martins}, {Schaerer}, \& {Hillier}}]{okali}
{Martins}, F., {Schaerer}, D., \& {Hillier}, D.~J. 2005, \aap, 436, 1049

\bibitem[{{Naz{\'e}} {et~al.}(2013){Naz{\'e}}, {Oskinova}, \& {Gosset}}]{nazog}
{Naz{\'e}}, Y., {Oskinova}, L.~M., \& {Gosset}, E. 2013, \apj, 763, 143

\bibitem[{{Norman} {et~al.}(1980){Norman}, {Wilson}, \& {Barton}}]{norwi}
{Norman}, M.~L., {Wilson}, J.~R., \& {Barton}, R.~T. 1980, \apj, 239, 968

\bibitem[{{Oskinova} {et~al.}(2007){Oskinova}, {Hamann}, \&
  {Feldmeier}}]{osporcar}
{Oskinova}, L.~M., {Hamann}, W.-R., \& {Feldmeier}, A. 2007, \aap, 476, 1331

\bibitem[{{Owocki}(1991)}]{owosam}
{Owocki}, S.~P. 1991, in Wolf-Rayet Stars and Interrelations with Other Massive
  Stars in Galaxies, ed. K.~A. {van der Hucht} \& B.~{Hidayat}, Vol. 143, 155

\bibitem[{{Owocki} {et~al.}(1988){Owocki}, {Castor}, \& {Rybicki}}]{ocr}
{Owocki}, S.~P., {Castor}, J.~I., \& {Rybicki}, G.~B. 1988, \apj, 335, 914

\bibitem[{{Owocki} \& {Puls}(1996)}]{owpu0}
{Owocki}, S.~P. \& {Puls}, J. 1996, \apj, 462, 894

\bibitem[{{Owocki} \& {Puls}(1999)}]{owpu}
{Owocki}, S.~P. \& {Puls}, J. 1999, \apj, 510, 355

\bibitem[{{Owocki} \& {Rybicki}(1984)}]{ornest}
{Owocki}, S.~P. \& {Rybicki}, G.~B. 1984, \apj, 284, 337

\bibitem[{{Pedersen} {et~al.}(2019){Pedersen}, {Chowdhury}, {Johnston},
  {Bowman}, {Aerts}, {Handler}, {De Cat}, {Neiner}, {David-Uraz}, {Buzasi},
  {Tkachenko}, {Sim{\'o}n-D{\'\i}az}, {Moravveji}, {Sikora}, {Mirouh},
  {Lovekin}, {Cantiello}, {Daszy{\'n}ska-Daszkiewicz}, {Pigulski},
  {Vanderspek}, \& {Ricker}}]{dalsipedersen}
{Pedersen}, M.~G., {Chowdhury}, S., {Johnston}, C., {et~al.} 2019, \apjl, 872,
  L9

\bibitem[{{Poe} {et~al.}(1990){Poe}, {Owocki}, \& {Castor}}]{havran}
{Poe}, C.~H., {Owocki}, S.~P., \& {Castor}, J.~I. 1990, \apj, 358, 199

\bibitem[{{Ramiaramanantsoa} {et~al.}(2018){Ramiaramanantsoa}, {Moffat},
  {Harmon}, {Ignace}, {St-Louis}, {Vanbeveren}, {Shenar}, {Pablo},
  {Richardson}, {Howarth}, {Stevens}, {Piaulet}, {St-Jean}, {Eversberg},
  {Pigulski}, {Popowicz}, {Kuschnig}, {Zoc{\l}o{\'n}ska}, {Buysschaert}, {Hand
  ler}, {Weiss}, {Wade}, {Rucinski}, {Zwintz}, {Luckas}, {Heathcote},
  {Cacella}, {Powles}, {Locke}, {Bohlsen}, {Chen{\'e}}, {Miszalski}, {Waldron},
  {Kotze}, {Kotze}, \& {B{\"o}hm}}]{ram}
{Ramiaramanantsoa}, T., {Moffat}, A. F.~J., {Harmon}, R., {et~al.} 2018,
  \mnras, 473, 5532

\bibitem[{{Reeve} \& {Howarth}(2016)}]{okraaj}
{Reeve}, D.~C. \& {Howarth}, I.~D. 2016, \mnras, 456, 1294

\bibitem[{{Ricker} {et~al.}(2015){Ricker}, {Winn}, {Vanderspek}, {Latham},
  {Bakos}, {Bean}, {Berta-Thompson}, {Brown}, {Buchhave}, {Butler}, {Butler},
  {Chaplin}, {Charbonneau}, {Christensen-Dalsgaard}, {Clampin}, {Deming},
  {Doty}, {De Lee}, {Dressing}, {Dunham}, {Endl}, {Fressin}, {Ge}, {Henning},
  {Holman}, {Howard}, {Ida}, {Jenkins}, {Jernigan}, {Johnson}, {Kaltenegger},
  {Kawai}, {Kjeldsen}, {Laughlin}, {Levine}, {Lin}, {Lissauer}, {MacQueen},
  {Marcy}, {McCullough}, {Morton}, {Narita}, {Paegert}, {Palle}, {Pepe},
  {Pepper}, {Quirrenbach}, {Rinehart}, {Sasselov}, {Sato}, {Seager},
  {Sozzetti}, {Stassun}, {Sullivan}, {Szentgyorgyi}, {Torres}, {Udry}, \&
  {Villasenor}}]{commander}
{Ricker}, G.~R., {Winn}, J.~N., {Vanderspek}, R., {et~al.} 2015, Journal of
  Astronomical Telescopes, Instruments, and Systems, 1, 014003

\bibitem[{{Sim{\'o}n-D{\'\i}az} {et~al.}(2018){Sim{\'o}n-D{\'\i}az}, {Aerts},
  {Urbaneja}, {Camacho}, {Antoci}, {Fredslund Andersen}, {Grundahl}, \&
  {Pall{\'e}}}]{simondia}
{Sim{\'o}n-D{\'\i}az}, S., {Aerts}, C., {Urbaneja}, M.~A., {et~al.} 2018, \aap,
  612, A40

\bibitem[{{Stone} \& {Norman}(1992)}]{zeus}
{Stone}, J.~M. \& {Norman}, M.~L. 1992, \apjs, 80, 753

\bibitem[{{Sundqvist} \& {Owocki}(2015)}]{sunowroz}
{Sundqvist}, J.~O. \& {Owocki}, S.~P. 2015, \mnras, 453, 3428

\bibitem[{{Sundqvist} {et~al.}(2018){Sundqvist}, {Owocki}, \& {Puls}}]{sundsim}
{Sundqvist}, J.~O., {Owocki}, S.~P., \& {Puls}, J. 2018, \aap, 611, A17

\bibitem[{{ud-Doula} \& {Owocki}(2002)}]{udo}
{ud-Doula}, A. \& {Owocki}, S.~P. 2002, \apj, 576, 413

\bibitem[{{{\v S}urlan} {et~al.}(2013){{\v S}urlan}, {Hamann}, {Aret},
  {Kub{\'a}t}, {Oskinova}, \& {Torres}}]{clres2}
{{\v S}urlan}, B., {Hamann}, W.-R., {Aret}, A., {et~al.} 2013, \aap, 559, A130

\bibitem[{{van Leer}(1977)}]{bram}
{van Leer}, B. 1977, Journal of Computational Physics, 23, 276

\end{thebibliography}

\end{document}